\documentclass{sig-alternate}

\usepackage{graphicx}
\usepackage{amsmath}   
\usepackage{mathtools}  
\usepackage{amsfonts}  
\usepackage{url}     
\usepackage{color}  
\usepackage{hyperref} 
\usepackage{threeparttable} 
\usepackage{amsmath, amsfonts, amssymb} 
\usepackage{parsetree} 
\usepackage{multicol} 

\begin{document}
%
\conferenceinfo{Cyber and Information Security Research Conference}{2015 Oak Ridge, TN}
\CopyrightYear{2015} 
\crdata{978-1-4503-3345-0/15/04\\
\url{http://dx.doi.org/10.1145/2746266.2746277}}  


\title{
Towards a Relation Extraction Framework for Cyber-Security Concepts 
\titlenote{\scriptsize{This manuscript has been authored by UT-Battelle, LLC under Contract No. DE-AC05-00OR22725 with the U.S. Department of Energy.  The United States Government retains and the publisher, by accepting the article for publication, acknowledges that the United States Government retains a non-exclusive, paid-up, irrevocable, world-wide license to publish or reproduce the published form of this manuscript, or allow others to do so, for United States Government purposes.  The Department of Energy will provide public access to these results of federally sponsored research in accordance with the DOE Public Access Plan \url{http://energy.gov/downloads/doe-public-access-plan}. 
}}
}

%
%
%
%
%

\numberofauthors{1}
\author{
\end{tabular}
\begin{tabular}{cccc}	
	Corinne L. Jones & Robert A. Bridges & Kelly M. T. Huffer & John R. Goodall\\
		\email{\href{mailto:cjones6@uw.edu}{cjones6@uw.edu}}
 & \email{\href{mailto:brigesra@ornl.gov}{bridgesra@ornl.gov}} &\email{\href{mailto:testakm@ornl.gov}{testakm@ornl.gov}} &\email{\href{mailto:jgoodall@ornl.gov}{jgoodall@ornl.gov}}
\end{tabular}
\begin{tabular}{c}
		\affaddr{University of Washington}\\	
		\affaddr{Department of Statistics}\\
		\affaddr{Box 354322}\\
		\affaddr{Seattle, WA 98195}
\end{tabular}
\begin{tabular}{c}
	\affaddr{Cyber \& Information Security Research Group}\\
	\affaddr{Computational Sciences and Engineering Division}\\
	\affaddr{Oak Ridge National Laboratory}\\
    \affaddr{Oak Ridge, TN 37831}
}

%

\date{\today}

\maketitle
\begin{abstract}
In order to assist security analysts in obtaining information pertaining to their network, such as novel vulnerabilities, exploits, or patches, information retrieval methods tailored to the security domain are needed. 
As labeled text data is scarce and expensive, we follow developments in semi-supervised Natural Language Processing and implement a bootstrapping algorithm for extracting security entities and their relationships from text. 
The algorithm requires little input data, specifically, a few relations or patterns (heuristics for identifying relations), and incorporates an active learning component which queries the user on the most important decisions to prevent drifting from the desired relations. 
Preliminary testing on a small corpus shows promising results, obtaining precision of .82. 
\end{abstract}

\section{Introduction}
The overall motivation behind this work is to aid security analysts in finding and understanding information on  vulnerabilities, attacks, or patches that are applicable to their network.  
As discussed in \cite{mcneil2013pace}, public disclosure of vulnerabilities and exploits often occur in obscure text sources, such as forums or mailing lists, sometimes months before inclusion in databases such as the CVE or NVD.\footnote{\url{https://cve.mitre.org/}, \url{https://nvd.nist.gov/}} 
Hence, tailored methods of automated information retrieval are needed for immediate awareness of security flaws. 
Additionally, our experience with analysts shows that necessary tasks, such as triage and response to alerts, performing network forensics, and seeking mitigation techniques, require security practitioners to find and process a large amount of information that is external to, but applicable to their network. 
Such information resides in online resources such as vulnerability and exploit databases, vendor bulletins, news feeds,  and security blogs, and is either unstructured text or a structured database with text description fields. 
In short, there is a critical need to aid security analysts in processing text-based sources by providing automated information retrieval and organization techniques. 
To address this, we seek entity and relation extraction techniques to identify software vendors, products, and versions in text along with 
vulnerability terms and the mutual associations among these entities. 

Relation extraction is the area of natural language processing (NLP) that seeks to recover structured data in the form of \textit{(subject entity, predicate relation, object entity)}-triples that match a database schema from text sources. 
For example, from the sentence ``Microsoft has released a fix for a critical bug that affected its Internet Explorer browser.'', we would like to extract (Microsoft, is\_vendor\_of, Internet Explorer) as an instance of the (software vendor, is\_vendor\_of, software product) relation. 
Our choices for entity concepts and their mutual relations are driven by an ontology of security domain, which gives a schema for storage in a graph database (see Sections \ref{entity-extraction}, \ref{patterns}) ideally making such information easily accessible to analysts. 
As annotated textual data is scant and costly to produce, we describe a semi-supervised technique for entity and relation extraction that incorporates active learning (querying the user) to assist in labeling only a few, most influential instances. 
While our implementation combines elements of many previous bootstrapping systems in the literature, applying these methods to relation extraction in cyber-security is to the best of our knowledge novel. 
Preliminary testing on a small corpus gives promising results; in particular, low false positive rates are obtained. 
Ideally, this preliminary work will lead to a fully developed operational version that automatically populates a knowledge base of cyber-security concepts from pertinent text sources.

\section{Bootstrapping}
As is often the case with deploying NLP techniques to a specific domain, the main hurdle is the lack of annotated data, which is expensive to produce. 
Consequently, supervised means of information extraction, although thoroughly developed, are not applicable. 
To accommodate this constraint, we implement a semi-supervised approach for relationship extraction that follows the previous work in the literature, but is tailored to our needs.
Our implementation builds on Brin's Dual Iterative Pattern Relation Expansion DIPRE algorithm \cite{brin1998extracting}, which uses a cyclic process to iteratively build known relation instances and heuristics for finding those instances.   
Input to the algorithm is (1) a relatively small set of ``seed'' instances of a given relation and/or seed patterns (heuristics for identifying a relation instance generally from the surrounding text) and (2) a relatively large corpus of unlabeled documents that are believed to contain many instances of the relation. 
The process proceeds by searching the corpus for mentions of the few known seed instances, and, upon finding an instance, the system automatically generates patterns from the surrounding text and stores these heuristics. 
Next, the corpus is traversed a second time using the patterns to identify any, hopefully new, instances of the relation, which are then stored in the seed set. 
To prevent the system from straying to undesired relations and patterns, a common addition to the DIPRE process is a method of confidence scoring for nominated patterns/relation instances \cite{carlson2010active}.
Our scoring procedure follows the BASILISK method of Thelen \& Riloff~\cite{thelen2002bootstrapping} and is discussed in Section \ref{scoring}. 
A variety of works have made contributions to this basic bootstrapping process, and we reference the improvements that have shaped our implementation throughout the discussion. 
We note that this bootstrapping process has been used solely for entity extraction in many works, in particular, \cite{carlson2010toward, carlson2010active, etzioni2005unsupervised, mcneil2013pace, thelen2002bootstrapping}. 
 

\subsection{Entity Extraction \& Document Relevancy}
\label{entity-extraction}
In order to extract a relation instance, the two entities involved must be identified, and this task has previously been accomplished in two ways. 
First, by using relation extraction patterns that identify entities; for instance, the initial work by Brin used regular expressions as part of the patterns to identify potential entities and relations simultaneously \cite{brin1998extracting}. 
Secondly, a common technique for relation extraction algorithms (bootstrapping and otherwise) is by using named entity recognition (NER) tools to identify entities, and afterward proceed with techniques to find entity pairs which share the desired relation \cite{agichtein2000snowball, alfonseca2012pattern, de2013unsupervised, mintz2009distant, more2012knowledge, mulwad2011extracting}.
 Unfortunately, our experience as well as that of previous works have found that ``off-the-shelf'' NER tools often fail to identify many cyber-security domain concepts \cite{bridges2014automatic,  joshi2013extracting, mcneil2013pace}. 
Using Freebase,\footnote{\url{http://www.freebase.com/}} an online knowledge base, for entity extraction and disambiguation occurs in previous works \cite{alfonseca2012pattern, mintz2009distant}, 
and more generally, calling on public databases to aid information retrieval is present in related works \cite{bridges2014automatic, more2012knowledge, mulwad2011extracting}. 

\begin{table}
\centering
\caption{Entity Types \& Extraction Methods}
\begin{tabular}{c|c|c}
\label{entity-table}
Entity Type & Example(s) & Extraction Method 	\\
\hline
SW\_Vendor	& Adobe 	&  Gazetteer (Freebase)\\  	
SW\_Product 	& Acrobat 		&  Gazetteer (Freebase) \\ 
SW\_Version	& 7, 11.0.08 		& Regular Expressions\\
CVE\_ID 		& CVE-2014-1127		& Regular Expression\\
MS\_ID		& MS-14-011			& Regular Expression\\
Vuln\_Term 	& xss, sql injection & Gazetteer (handmade)\\
SW\_Symbol 	& pAlloc(), reg.exe & Regular Expressions
\end{tabular}
\begin{tablenotes}
\small
\item Note: SW\_Product includes operating systems and applications. Vuln\_Term entities are members of a hand-crafted list developed in \cite{bridges2014automatic} that are terms descriptive of vulnerabilities or attack consequences and are similar to the ``Attack'' entity type in \cite{joshi2013extracting}. SW\_Symbol includes file names and named elements of code such as functions, methods, and classes.
\end{tablenotes}
\end{table}

Following this line of thought, our implementation uses gazetteers and regular expressions to label entities and discards documents deemed irrelevant before the bootstrapping begins.  
Entity types with examples and extraction method are detailed in Table \ref{entity-table}.
Lists of software vendors and software products were queried from the Freebase API under the \textit{/computer/software\_developer} category and \textit{/computer/software}, \textit{/computer/software/operating\_system}, re-\\spectively.
 A major advantage of this approach is that Freebase includes alternate names for entities allowing easy disambiguation;  e.g., ``IE'' and  ``Internet Explorer'' are all identified as aliases of the same entity in Freebase's database. 
In order to discard documents unlikely to contain a relation before further processing, a logistic regression classifier is trained on a few documents hand-labeled as relevant or not. 
For our implementation, the number of each entity type appearing in the document are the input features, although the framework is designed for a more robust feature set if desired. 
Lastly, during the bootstrapping process for relation extraction, if a known pattern is identified with an unlabeled entity, that entity is then labeled. 
Hence, in the learning of relations and patterns, the bootstrapping process can aid in identification of entities omitted by the initial gazetteers or regular expression taggers.

%

\subsection{Relations and Patterns}
\label{patterns}
The bootstrapping process is run independently for each of 9 relation types listed in Table \ref{relation-table}.
Informed by research efforts to build an ontology of the security domain \cite{iannacone2015developing},
the first eight relations correspond to attributes of the ``Vulnerability'' and ``Software'' nodes of the ontology, while the last relation, ``not\_version\_of'' was added to indicate and prevent common errant inferences of the opposite relation. 
For each relation, a few hand-crafted rules are given as input to start the bootstrapping process, as reported in Table \ref{relation-table}. 
\begin{table}
\centering
\caption{Relation Types}
\begin{tabular}{c|c|c|c|c}
\label{relation-table}
	& Subject Entity 	& Relation 			& Object Entity		& N\\
\hline
1	& SW\_Vendor			& is\_vendor\_of		& SW\_Product		& 14 \\
2 	& SW\_Version		& is\_version\_of	& SW\_Product		& 6\\
3	& CVE\_ID			& CVE\_of\_vuln		& Vuln\_term			& 1\\
4	& MS\_ID				& MS\_of\_SW			& SW\_Product		& 6\\		
5	& MS\_ID 			& MS\_of\_vuln		& Vuln\_term			& 7\\
6	& Vuln\_term			& vuln\_of\_SW	& SW\_Product		& 14\\
7	& SW\_Symbol			& symbol\_of 	& SW\_Product	& 2\\
8	& SW\_Version		& not\_version\_of	& SW\_Product		& 9\\
\end{tabular}
\begin{tablenotes}
\small
\item Note: N denotes number of seed patterns for each relation.  
\end{tablenotes}
\end{table}

Our implementation involves three types of patterns: 
(1) all the words/parts-of-speech in order between two appropriate entity types, 
(2) a subset of contiguous words/parts-of-speech occurring between two appropriate entity types,
(3) the parse tree path between two appropriate entity types.  
 A parse tree is an assignment of a tree structure to each sentence with the words as leaf nodes, words' parts-of-speech as their parents, and other ancestral nodes correspond to the sentence's structure as given by a context-free grammar. 
 By definition there exists a unique path between any two nodes in a tree, and, in particular, for two known entities in a sentence, the parse tree path, known as a dependency path, gives a natural feature set for deducing the existence of a relation between the entities \cite{alfonseca2012pattern, bunescu2005shortest, de2013unsupervised, mintz2009distant}. 
\begin{multicols}{2}
\phantom{blah\\}
\noindent Example parse tree for sentence ``I like eggs.'' is given. For this example the pattern we extract from the parse tree path from ``I'' to ``eggs'' is [N, NP, S, VP, N]. 
\columnbreak
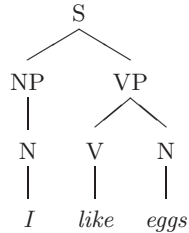
\begin{parsetree}
	( .S. 
        (.NP.  
         	(.N. `I')
		)        
        (.VP.
            (.V. `like' )
            (.N.  `eggs')
        )
	)
\end{parsetree}
\end{multicols}




\subsection{Scoring \& Active Learning}
\label{scoring}
A common pitfall of bootstrapping algorithms is straying from the desired topic by iteratively learning spurious patterns and relations. 
To increase accuracy, a variety of scoring methods for rating  confidence in a nominated relation or pattern have been proposed  \cite{carlson2010active, etzioni2005unsupervised, jones1999bootstrapping}, and the common, underlying goals of scoring methods are  (1) to seek patterns that are indicative of the given relation, but do not occur with spurious instances, and (2) to seek relations that only occur with trusted patterns. 
Additionally, active learning, which refers to systems that query the user to provide accurate input on a few pertinent examples, is incorporated into our system, and has been used in a few previous bootstrapping works \cite{carlson2010active, thelen2002bootstrapping}. 
Discussion and comparative evaluation of five bootstrapping scoring methods, in particular in the presence of active learning is the topic of \cite{carlson2010active}. 
As the BASILISK system of \cite{thelen2002bootstrapping} showed the greatest benefit from user interaction in the study, our scoring procedure is inherited from their implementation as described here \cite{jones1999bootstrapping}.
 
Specifically, if a potential relation instance, $r$, is identified by distinct patterns $p_1, ..., p_n$, R\_score$(r):= \sum_{i=1}^n \log(f_i+1) /n$ where $f_i$ is the number of unique known relations identified by $p_i$. 
Thus, relations identified by many successful patterns will score the highest. 
If a potential pattern $p$ is nominated by occurring at least once with unique known relations $r_1, \dots, r_m$, then P\_score$(p):= m\log(m)/N$ where $N=$ number of unique occurrences of $p$ (with or without a known relation). 
Hence, the number of known instances the pattern matches is weighted by the pattern's precision, and patterns that both match many known relations and have high precision obtain the highest scores.  
To help prevent the system from drifting, an option for user interaction is incorporated by allowing specification of the number of queries per cycle. 
We note that the highest scoring patterns are the ones that nominated the most relations, and therefore have the largest effect on the direction of the system; 
thus, for relations and patterns scoring the highest, the system asks a user to verify their validity with ``yes'', ``no'', ``don't know'' response options. 
Given a response of ``yes'' (``no''), the score is set to 1000 (-1). 
After the scoring, relations/patterns with scores in a sufficiently high percentile are added to the set of relation/pattern seeds for the next cycle.
In the case of conflicting relation nominations (e.g., ``is\_version\_of'' and ``not\_version\_of'') the system queries the user if possible and defaults first to a heuristic in some cases and then to the highest scoring relation. 
Lastly, when a known pattern is found in the text with an unlabeled entity, the system queries the user if possible before labeling the entity.

\section{Results}
\label{results}
For initial testing a corpus of 62 news articles, blogs, and updates is compiled from a variety of security-related websites.\footnote{Text sources from \url{www.arstechnica.com},  \url{www.computerworld.com}, \url{www.krebsonsecurity.com}, \url{www.schneier.com}, \url{www.security.blogoverflow.com}, \url{www.securityintelligence.com}, \url{www.smh.com.au}, \url{http://technet.microsoft.com}, \url{http://blog.cmpxchg8b.com}, \url{https://blog.malwarebytes.org}, \& \url{www.threatpost.com}.}
After pulling each site's text with the Goose\footnote{\url{https://pypi.python.org/pypi/goose-extractor/}} article extractor, word-, sentence-tokenization, and part-of-speech tagging is performed using CoreNLP~\cite{manningstanford}. 
Only after the entity extractor provides entity labels and the relevancy classifier discards irrelevant documents (see Section \ref{entity-extraction}) are parse trees applied by CoreNLP to the remaining documents\textemdash 41 
documents were kept, parsed, and used as the corpus for bootstrapping. 
With an eye on scaling to a large corpus, this is a noteworthy detail as the parsing is computationally and temporally expensive (e.g., parsing took about 300 of 370 seconds for the NLP pipeline).   

As no labeled corpus of relations for our domain was available, evaluation is difficult; in particular, calculating recall would involve labeling all the documents. 
To get a rough estimate of the recall of the system, we hand labeled a longer article, obtaining 8 correctly identified of 33 total relations.  
Hence at least locally the recall is 0.24.   
By manually checking each relation instance the system output, we provide precision results in Table \ref{results-table}. 
The reported run accepted the top 80\% of patterns and relations after each iteration of scoring and queried the user for the top 2\%, resulting in $\sim5$ queries per iteration. 
Scores are reported for output after 3 iterations through the corpus, after which point the number of relations did not increase. 
Of the 41 relevant documents, 31 included an identified relation, and 186 overall relations were identified of which 153 are correct. 
We note that the seed patterns were crafted from example sentences observed by the authors, and, hence, testing on a larger corpus will be needed to flatten any bias that may have come from our observation of some of the 62 documents. 

\begin{table}
\label{results-table}
\centering
\caption{Results}
\begin{tabular}{c|c|c|c|c}
	& Relation                          & \textbf{TP} & \textbf{FP} & \textbf{P} \\
\hline
1 & is\_vendor\_of           & 45          & 12          & 0.79       \\
2 & is\_version\_of              & 54          & 19          & 0.74       \\
3 & CVE\_of\_vuln & -		& - & - \\
4 & MS\_of\_SW	 & - & - & -\\
5 & MS\_of\_vuln	  & 2           & 0           & 1.00       \\
6 & vuln\_of\_SW & 30          & 2           & 0.94       \\
7 & symbol\_of 	 &  - & - & -\\
8 & not\_version\_of        & 22          & 0           & 1.00       \\
\hline
& \textbf{Totals}                   & \textbf{153}         & \textbf{33}         & \textbf{0.82}      
\end{tabular}
\begin{tablenotes}
\small
\item Note: True positives, false positives, and precision by relation type and in total reported. Dashes indicate no instances were found in the corpus.
\end{tablenotes}
\end{table}

\section{Conclusion}
In summary, our bootstrapping algorithm is a promising start to a pertinent problem, namely, the need for automated information extraction targeting security documentation.  
As our preliminary tests involved a relatively small corpus, further testing on a larger corpus is necessary. 
Additionally, comparative evaluation to quantify the benefit of incorporating active learning is a desired future direction. 
Ultimately, we plan to incorporate this work into a larger pipeline, which continually feeds it new documents from the web, and organizes the output into a database.  
\section{Acknowledgments}
The authors thank Jason Laska and Michael Iannacone for helpful discussions. 
This material is based on research sponsored by the Department of Homeland Security Science and Technology Directorate, Cyber Security Division via BAA 11-02; the Department of National Defence of Canada, Defence Research and Development Canada; the Dutch Ministry of Security and Justice; and the Department of Energy. 
The views and conclusions contained herein are those of the authors and should not be interpreted as necessarily representing the official policies or endorsements, either expressed or implied, of: the Department of Homeland Security; the Department of Energy; the U.S. Government; the Department of National Defence of Canada, Defence Research and Development Canada; or the Dutch Ministry of Security and Justice.


\bibliographystyle{abbrv}
\bibliography{rel-ext} 

\begin{thebibliography}{10}

\bibitem{agichtein2000snowball}
E.~Agichtein and L.~Gravano.
\newblock Snowball: Extracting relations from large plain-text collections.
\newblock In {\em Proceedings of the Fifth ACM Conference on Digital
  Libraries}, pages 85--94. ACM, 2000.

\bibitem{alfonseca2012pattern}
E.~Alfonseca, K.~Filippova, J.-Y. Delort, and G.~Garrido.
\newblock Pattern learning for relation extraction with a hierarchical topic
  model.
\newblock In {\em Proceedings of the 50th Annual Meeting of the Association for
  Computational Linguistics: Short Papers-Volume 2}, pages 54--59. Association
  for Computational Linguistics, 2012.

\bibitem{bridges2014automatic}
R.~A. Bridges, C.~Jones, M.~Iannacone, K.~Testa, and J.~R. Goodall.
\newblock Automatic labeling for entity extraction in cyber security.
\newblock In {\em Third Annual ASE Cyber Security Conference}. {ASE}, 2014.

\bibitem{brin1998extracting}
S.~Brin.
\newblock Extracting patterns and relations from the world wide web.
\newblock In {\em Selected Papers from the International Workshop on The World
  Wide Web and Databases}, WebDB '98, pages 172--183, London, UK, UK, 1999.
  Springer-Verlag.

\bibitem{bunescu2005shortest}
R.~C. Bunescu and R.~J. Mooney.
\newblock A shortest path dependency kernel for relation extraction.
\newblock In {\em Proceedings of the conference on Human Language Technology
  and Empirical Methods in Natural Language Processing}, pages 724--731.
  Association for Computational Linguistics, 2005.

\bibitem{carlson2010toward}
A.~Carlson, J.~Betteridge, B.~Kisiel, B.~Settles, E.~R. Hruschka~Jr, and T.~M.
  Mitchell.
\newblock Toward an architecture for never-ending language learning.
\newblock In {\em AAAI}, volume~5, page~3, 2010.

\bibitem{carlson2010active}
A.~Carlson, S.~A. Hong, K.~Killourhy, and S.~Wang.
\newblock Active learning for information extraction via bootstrapping, 2010.

\bibitem{de2013unsupervised}
O.~L. de~Lacalle and M.~Lapata.
\newblock Unsupervised relation extraction with general domain knowledge.
\newblock In {\em EMNLP}, pages 415--425, 2013.

\bibitem{etzioni2005unsupervised}
O.~Etzioni, M.~Cafarella, D.~Downey, A.-M. Popescu, T.~Shaked, S.~Soderland,
  D.~S. Weld, and A.~Yates.
\newblock Unsupervised named-entity extraction from the web: An experimental
  study.
\newblock {\em Artificial Intelligence}, 165(1):91--134, 2005.

\bibitem{iannacone2015developing}
M.~D. Iannacone, S.~Bohn, G.~Nakamura, J.~Gerth, K.~M.~T. Huffer, R.~A.
  Bridges, E.~Ferragut, and J.~T. Goodall.
\newblock Developing an ontology for cyber security knowledge graphs.
\newblock In {\em Proceedings of the {CISRC-10}, the Tenth Cyber \& Information
  Security Research Conference}. ACM, 2015.

\bibitem{jones1999bootstrapping}
R.~Jones, A.~McCallum, K.~Nigam, and E.~Riloff.
\newblock Bootstrapping for text learning tasks.
\newblock In {\em IJCAI-99 Workshop on Text Mining: Foundations, Techniques and
  Applications}, volume~1. Citeseer, 1999.

\bibitem{joshi2013extracting}
A.~Joshi, R.~Lal, T.~Finin, and A.~Joshi.
\newblock Extracting cybersecurity related linked data from text.
\newblock In {\em {IEEE} Seventh International Conference on Semantic Computing
  ({ICSC})}, pages 252--259. {IEEE}, 2013.

\bibitem{manningstanford}
C.~D. Manning, M.~Surdeanu, J.~Bauer, J.~Finkel, S.~J. Bethard, and
  D.~McClosky.
\newblock The {Stanford CoreNLP Natural Language Processing Toolkit}.

\bibitem{mcneil2013pace}
N.~McNeil, R.~A. Bridges, M.~D. Iannacone, B.~Czejdo, N.~Perez, and J.~R.
  Goodall.
\newblock {PACE}: Pattern accurate computationally efficient bootstrapping for
  timely discovery of cyber-security concepts.
\newblock In {\em 2013 12th International Conference on Machine Learning and
  Applications ({ICMLA})}, volume~2, pages 60--65. IEEE, 2013.

\bibitem{mintz2009distant}
M.~Mintz, S.~Bills, R.~Snow, and D.~Jurafsky.
\newblock Distant supervision for relation extraction without labeled data.
\newblock In {\em Proceedings of the Joint Conference of the 47th Annual
  Meeting of the ACL and the 4th International Joint Conference on Natural
  Language Processing of the AFNLP}, volume~2, pages 1003--1011. Association
  for Computational Linguistics, 2009.

\bibitem{more2012knowledge}
S.~More, M.~Matthews, A.~Joshi, and T.~Finin.
\newblock A knowledge-based approach to intrusion detection modeling.
\newblock In {\em Security and Privacy Workshops (SPW), 2012 IEEE Symposium
  on}, pages 75--81. IEEE, 2012.

\bibitem{mulwad2011extracting}
V.~Mulwad, W.~Li, A.~Joshi, T.~Finin, and K.~Viswanathan.
\newblock Extracting information about security vulnerabilities from web text.
\newblock In {\em Web Intelligence and Intelligent Agent Technology Conference
  (WI-IAT), IEEE/WIC/ACM}, volume~3, pages 257--260. IEEE, 2011.

\bibitem{thelen2002bootstrapping}
M.~Thelen and E.~Riloff.
\newblock A bootstrapping method for learning semantic lexicons using
  extraction pattern contexts.
\newblock In {\em Proceedings of the ACL conference on Empirical Methods in
  {NLP}}, volume~10, pages 214--221. Association for Computational Linguistics,
  2002.

\end{thebibliography}
\end{document}